# Magic Fairy Tales as Source for Interface Metaphors

Vladimir L. Averbukh
Institute of Mathematics and Mechanics UrB RAS
Ekaterinburg Russia
averbukh@imm.uran.ru

## ABSTRACT
The work is devoted to a problem of search of metaphors for interactive systems and systems based on Virtual Reality (VR) environments. The analysis of magic fairy tales as a source of metaphors for interface and virtual reality is offered. Some results of design process based on "magic" metaphors are considered.

## Categories and Subject Descriptors
K.m MISCELLANEOUS

## General Terms
Human Factors

## Keywords
Computer metaphor, interface

## 1. INTRODUCTION
This work is devoted to a problem of search of metaphors for interactive systems and systems based on Virtual Reality (VR) environments. Metaphors play a significant role while designing, and also help users to master interface opportunities with ease. Sometimes the success or failure of system depends on successful choice of metaphors. However researchers have started to complain of rarity of interface metaphors already in early 90-s' years when development of visual interactive systems were anywise a novelty. (See [1].) In our opinion one may find a lot of sources of metaphors, but it is necessary to search these sources by some criteria. We suggest to consider magic fairy tales and popular fantasy novels (using a set of characters, concepts and situations similar to the magic fairy tales) as one of sources of computer metaphors. Thus, we are interested with "magic" ideas i.e. ideas of miraculous actions, positions, things and events that are impossible in reality. These ideas are useful to create interface and visualization metaphors and to develop new techniques for dealing in virtual environments. Note that VR may demand wide diversity of magic metaphors, because in fact it is a real ("not magic") magic. Even superficial analysis shows that at times magic in fairy tales extends possibilities and features of realistic objects and characters and at other times fairy tales describe occurrences being in marked contrast to everyday routine. Both cases are interesting in connection with metaphors searching. Moreover VR also at times augments user's means, and at other times transfers users to new worlds.

Magic ideas described in fantasy novels or in literary processed folklore fairy tales are the most worked and consequently the most useful for our purposes.

Interface metaphor is considered as the basic idea of likening between interactive objects and model objects of the application domain. Its role is to promote the best understanding of semantics of interaction, and also to determine the visual representation of dialog objects and a set of user manipulations with them. A metaphor, considered as a basis of the sign system, underlies in a basis of a dialog language in its turn. User formulates the problem with the help of this language and achieves its solving from the computer. The metaphor helps to describe abstraction, structures understanding of new applied area, but also assigns dialog [visual] language objects.

We understand interface metaphors in broadened sense, not subdividing them onto metaphors, analogies, idioms or metonymies. These divisions are important for the metaphor theory, but are not necessary for our consideration. Interface metaphors may be considered as a special case of scientific metaphor used for generation of new or additional senses to understand new facts and phenomena. The understanding of an interface metaphor as an usage only everyday and well-known (often technical) realities is unsatisfactory and even wrong for several reasons [2], [3].

## 2. RELATED WORKS
The problems of metaphor design, as well as formalizations of their search and/or generations were repeatedly considered in the literature. The idea of automated creation of visual metaphors and corresponding visualization systems was proposed.

Interest to "magic" in connection with HCI and interface metaphors was showed in early 90-s' years. The significant attention has been given to "Magic Features" conception. This conception in the context of HCI was introduced for the first time by A. Kay. [4]. In the literature (also by A. Kay) some theoretical approaches to problems of HCI magic were considered and some examples were given (see [5-9]. "Magic Features" are considered as important for HCI. The authors of the works [6-8] emphasized especially that metaphors doesn't preclude magic. Obviously, if to consider metaphors not only as usage everyday things and everyday entities then magic may be a part of interface metaphors. The word "magic" is very popular as a part of metaphoric names for interface techniques – for example "magic lens", "magic mirror", "magic lancet", etc. In [10] a number of specific examples from stage magic is presented and application of its principles and techniques in human interface design is discussed. The article [11] is devoted to sources of metaphor for tangible user interfaces. Authors suppose magic and paranormal phenomena could be a fruitful place to look for new metaphors for tangible user interfaces. Also Voodoo magic is considered as an interesting idea for interfaces with virtual objects. In [12] Voodoo dolls technique is used as a two-handed interaction technique for manipulating objects at a distance in immersive virtual environments.

Ideas for interface metaphor design, linked with magic fairy tales, fantasy and science fiction novels, are described in a number of articles. In [5] for example the fantastic metaphor of flying carpet is mentioned, but in [13] the metaphor of the magic carpet is realized for instant moving in the virtual reality environment. In [14] and [15] usages of "magic wand" are described. Magic wand is considered as a manipulation metaphor to form an interface in systems with elements of virtual reality. Interesting ideas of wonder objects (for example, magic mirror) were proposed (and realized in prototype variants) for storytelling in the modern museum [16]. In [3] the information system using a City Metaphor is described. In this system magic/fantastic opportunities are used on regular basis. Among these opportunities there is the "Tunneling through space" presenting the typical adoption from Science Fiction. Also rooms with "magic windows" may be considered as a magic (science fiction) feature. "Magic window" is an interesting expansion of a well-known Information Wall metaphor.

On our opinion the metaphors adopted from magic fairy tales, science fiction and fantasy novels are more preferable than the metaphors constructed on the basis of complex and gloomy magic rituals and Voodoo techniques.

Two famous book of Russian scientist Vladimir Propp (Morphology of the Folk Tale and The Historical Roots of the Wonder Tale) are devoted to fairy tales. (See translation one of them into English in [17].) In this work the sets of fantastic situations, characters (among them - the hero, the false hero, the assistant, the wrecker) and the structure of magic fairy tales are explored in detail, and also their connection with the basic ancient myths is shown.

In fairy tales of different nations one can find very interesting sets of "magic features". In the following sections we consider, what is behind this conception and how magic may be useful when searching of metaphors for interactive systems and systems based on virtual reality environments.

## 3. MAGIC. WHAT IS THIS?

Here we consider fantastic magic, selecting samples that are fruitful for interface and virtual reality metaphors.

**Magic transport.** In fairy tales and fantastic novels one can find:
*A*. Teleportation - instant carry by means of verbal influence (spell) or by means of manipulations with any objects or uses of such devices as teleports;
*B*. Rather slow moving by means of magic transport. A magic transfer may take place both for subjects, and for (animated and inanimate) objects of magic.

**Magic navigation means** - for example, the milestone with magic legend or the magic clew, following to which one may reach up Fairland.

**Magic communications means.** An example - the magic mirror tuned on the interesting character for his/her protection or observation.

**Wonderful** (additional to normal) **opportunities on manipulations** with objects, processes and even natural phenomena. In some variants - superforce, invulnerability, etc. Generally speaking one can use the term **"magic power"** (or **"superpower"**). These manipulations and power may be carried out through spells and objective magic, and also through the universal manipulator - a **magic wand**.

One can set off the general class **of the magic objects** as objects possessing "**magic properties**"**.** Thus it is possible the magic expansion of usual functionality (seven-league boots), as well as attributing to object of additional, unusual in reality functions (for example Aladdin's Wonderful Lamp used for a call, activization and neutralization **of magic beings**). For a designation of the magic objects used as manipulators, transport and communication means one can use also the term **"magic tools"**.

**Magic transformations of objects**. As an example one can consider the construction of palaces by Jinnee in the Arabic fairy tale or transformation a pumpkin into the carriage for Cinderella.

Similarly to magic objects **magic subjects** may be set off, i.e. evil, good or neutral (with respect to heroes) magicians possessing superopportunities. (As a variant - magic anthropomorphic beings, for example, fairies, gnomes, trolls, jinn, etc.)

**Magic transformations of persons**. In fairy tales such transformations may be spontaneous, unexpected for the characters, or transformations may be results of magic actions (some magic spells or manipulations). These transformations can be carried out in view of sympathetic magic (that is magic based on some similarity).

## 4. WHAT DOES PRACTICE SHOW?

Our attention was involved with two "magic-fantastic" metaphors from the novels (and also the films) about Harry Potter. These are *Speaking and Moving Portraits* and *the Map* on which shifts of persons under observation are visible. In this case portraits of dead persons are the active objects. They may address to alive characters of the novel without requests, and even pass from one portrait framework to another to do a visit to each other. The map continuously traces and shows the site of an observable person.

There are a lot of active, anthropomorphic and speaking characters in fairy tales and in science fiction novels such as Golem, robots of K. Capek and I. Asimov, and so on. Similarly there are various variants of magic/fantastic means of navigation and searching both in fairy tales, and in a fantasy.

In these magic metaphors visualness is not so important then spontaneous activity inherent in generated objects and subjects. Spontaneous activity can be considered as means to imitate reasonable behavior. It is necessary also subjectness of characters (i.e. their existence independently of users) to imitate reasonable behavior. In the fantastic and popular scientific literature and films such active computer "subjects" have appeared some decade ago. In modern computer practice agents who are active under own initiative, frequently cause irritation. We have been started our research of active intellectual agents to understand what, why and where have to do active intellectual agents.

The idea of the "active" map showing a real landscape and movings of objects on it was considered. "Activity" of a map can be connected to events, the same type, as well as in navigating systems - travel of this or that district, turn, crossing and so forth, but "activity" can be and "spontaneous", connected with time events. The other idea may be possible – the development of the

"active" scheme of a protected apartment or territory. In this case tracking systems and the "marked" persons may be necessary. Movings of all without exception persons may be shown on the scheme, and labels allow to identify them. In connection with the given ideas it is possible to note, that works in the given direction on the basis of systems such as GPS are actively carried out, and there are examples of the interactive maps which are performing roles of guide-adviser. Also now it is easy to develop speaking anthropomorphic agents-avatars, and there are a lot of examples of such realizations.

We have decided to connect the idea of active agents constructed on a metaphor of the speaking portrait, with the expert system. The point is that the active agent with its (possible) importunate activity is warranted almost only in case of learning systems. The logic of project development led us to the next idea the idea of "active textbook" [18]. This textbook have to be able to analyze the pupil behavior during textbook studying, for example, time of reading, manipulations with the text and so on. Basing on these analyses "active textbook" may detail teaching material, search new data sources or turn to other things. The analysis of a user behavior may be realized at a syntactic level (at a level of operations with the mouse and the keyboard, eye tracking, etc.), and on semantic one (monitoring of opened files or sites, applications under execution, recording of events, etc.) Such analyses and elements of programming by demonstrations will allow our system to learn to teach during the process of its using and to operate in the given direction "independently". Also it is possible to supply the system by means adviser functions. The system will be like clever human being – adviser and will not impose the opinion, but will prepare pieces of advice and decisions. The human-like behavior may be provide by psychologically caused braking of the system activity.

In the visualization systems constructed on the basis of virtual reality environments, there are the tasks where complex manipulations with objects are necessary, for example, to pull out something, to cut or to zoom. As a metaphor of the tool for such tasks, first of all the idea of a magic wand comes into designer's on mind. But a magic wand has not differentiated action and hence requires means to change modes of operating. In the specialized systems it is more natural to use specialized "magic tools". For example, in medical information system as manipulator's metaphor the idea of "magic lancet" is offered. "Lancet" allows "to dissects" this or that organism area for the profound exploring. During any human organism object "dissecting", all physical changes are visualized, as if we did it in a reality. In case of a combination the "magic lancet" metaphor with three-dimensional model of human body, one may obtain the virtual model of operations and the prototype system of information visualization for the medical purposes is under construction now. Systems based on this metaphor may be used for example for surgery learning [19].

## 5. CONCLUSIONS

Our preliminary research shows the applicability of "magic" metaphors for tasks in the interactive systems and systems basing on virtual reality environments. For example, the search of metaphors of moving in virtual environments may be needed magic transportation techniques, such as teleports and flights of various types (the flying carpet, the flying ship, **Roc**, a winged horse). And metaphors of intellectual agents-informants may be based on magic means of navigation.

In fairy tales and fantastic novels one can find a lot of the magic phenomena, such as **magic knowledge, war magic, fulfillments of desires, telepathy** and **thought-reading,** etc. But we yet don't know definitely, whether these feature useful for searching of metaphors. Though for a choice of metaphors for manipulations with objects and processes in virtual environments "**the war magic**" may be interesting. The **war magic** is connected with transferring the events which have place in magic space into reality. For example, any variations of "a magic chess", where games with chess-men are transferred into land battles, or the "naval" military magic where models of fight in a vat with water are transferred into sea battles. These ideas (partly close to Voodoo magic) one may find in a number of folk and literary fairy tales.

Let's pay attention that folk fairy tales are governed by rigorous logic of a plot development and a choice of characters. By the way, in literary fairy tales and fantasy novels, as a rule, this logic is observed.

The analysis shows, that very exotic "magic" metaphors may be useful to form any interface features. However realization of interactive systems on their base may be as complex as contradictory. Magic metaphors are frequently transformed to abstract interface opportunities, not keeping the appreciable connection with initial ideas. For example, in case of Speaking Portrait metaphor the anthropomorphness of an agent turned out not necessary. But there is necessary to endow it the function of the magic assistant, - conductor into the world of knowledge. Of course the transmuted abstractness of metaphors is an advantage, than a defect of their use. In the systems created for example for office automation or for end-user programming the presence of magic interface manuals may appear as distracting or even irritating factor. However use such "magic features" as automatic returning electronic analogs of paper documents on its place after the ending of processing (see [20]) may be carried out without any special warnings even for users - non-specialists. Such features are well conformed to common sense of clerks and does not demand unnecessary efforts during operations. Infringement of magic logic due to any absurd ideas or to farfetched subjective likeness may lead to serious mistakes. The sharp criticism of interface metaphors, as such, is connected to infringement of magic logic in early variant Apple's interface (using the trash can metaphor to eject disks) [4].